\documentclass[11pt]{article}

\def\leftdisplay$${\leftline{\noindent\ifleqno\eqn
  \indent\fi$\displaystyle{\eq}\ifeqno\hfill\ifleqno\else\eqn\fi\fi$}$$}
\newif\ifeqno \newif\ifleqno \everydisplay{\displaysetup}
\def\displaysetup#1$${\displaytest#1\eqno\eqno\displaytest}
\def\displaytest#1\eqno#2\eqno#3\displaytest{%
  \if!#3!\ldisplaytest#1\leqno\leqno\ldisplaytest
  \else\eqnotrue\leqnofalse\def\eqn{#2}\def\eq{#1}\fi\leftdisplay$$}
\def\ldisplaytest#1\leqno#2\leqno#3\ldisplaytest{\def\eq{#1}%
  \if!#3!\eqnofalse\else\eqnotrue\leqnotrue\def\eqn{#2}\fi}
\usepackage{graphicx}
\usepackage{nopageno}
\usepackage{lipsum}
 \def\ref{\par\noindent\hangindent=0.7 true cm
          \hangafter=1}
\def\h2{\hskip-2pt}
\begin{document}

 \Large
\centerline{\bf Differences between the values of frequencies by different fitters }
\vskip 4pt
\centerline{\bf 16CygA\&B and  Kepler Legacy values  } 
\large
\vskip 10pt
\centerline{Ian Roxburgh, Queen Mary University of London}
\normalsize
\vskip 12pt

\noindent The differences between the oscillation frequencies and uncertainy estimates of a star derived by different fitters can be large,  sufficiently large so that, were one to find a stellar model that fitted one frequency set ($\chi^2\sim 1$), it does not fit an alternative  set.  The table below gives 21 examples,  comparing  frequency sets in common  between the Kepler Legacy project  and  frequency sets from Appourchaux  et al (2014)$^1$ and Davies et al (2015)$^2$.  Figure 1 displays the frequency differences $\nu_L-\nu_D$ (Legacy-Davies) for 16CygA\&B and the $\chi^2$ of the fits to each other.  A model  whose frequencies fit the Legacy frequency set for 16CygA  with $\chi_L^2 <1$ could have $\chi_D^2>10$ for a fit to Davies's frequency set and so would be rejected.

%\noindent It can happen that the difference between the oscillation frequencies and errors of a star derived by different fitters can be large, sufficiently large so that were one to find a model that fitted one frequency set ($\chi^2\sim 1$) it may well not fit an alternative  set.  This is illustrated in the Table 1  below which compares frequency sets for 21 stars from the Kepler Legacy project with alternative frequency estimates from Appourchaux  et al (2014) and Davies et al (2015).  Prime examples are 16CygA\&B whose frequency differences, Legacy-Davies, and $\chi^2_m$ of the fit to each other  are shown below.  A model  whose frequencies fit the Legacy frequency set for 16CygA  with a $\chi_L^2 <1$ could have a $\chi_D^2>10$ for a fit to Davies's frequency set and would be rejected.
% \vskip-5pt
\begin{figure}[h]
\begin{center} 
\hskip-10pt
\includegraphics[ width=18.cm]{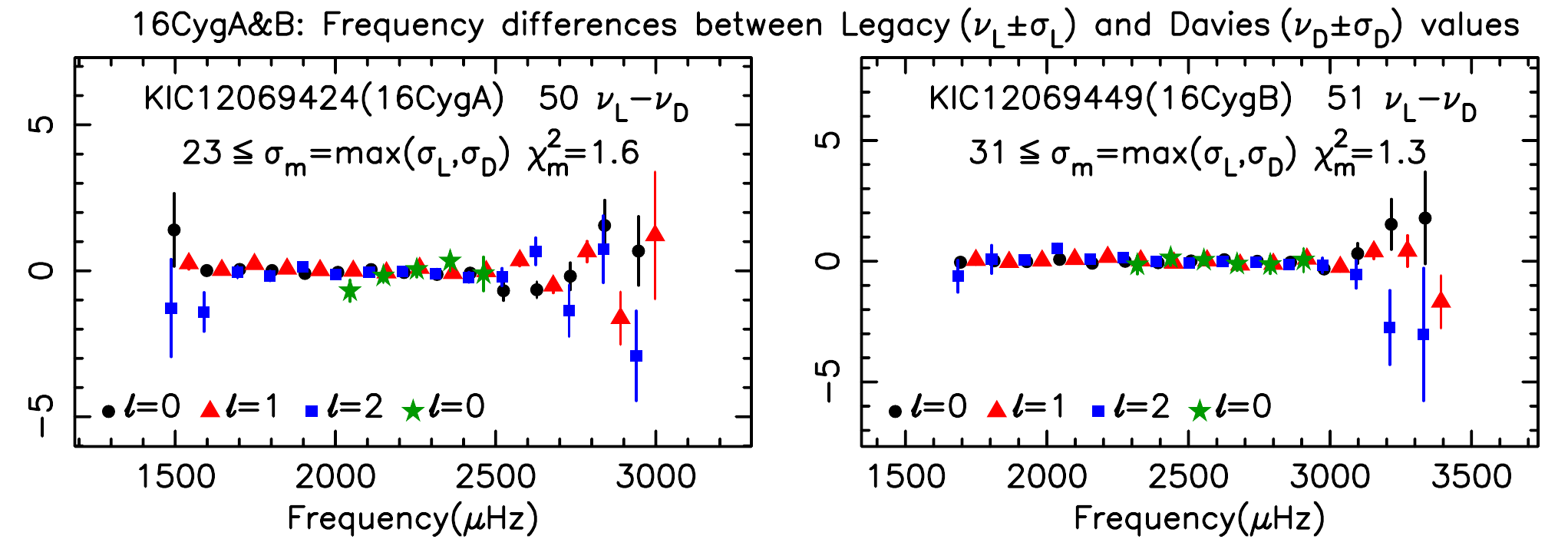}
  \end{center}  
 \end{figure}
\vskip-18pt
\noindent  These differences are not statistical uncertainties;  given the same input 
%data (an observed light curve) 
light curve,  differences in estimated frequencies are due to different assumptions/constraints in mode fitting techniques, the segment of the time series used, and the algorithms for determining power spectra.
% from the light curve. 
The differences constitute uncertainties in the values of the frequencies and should be added to estimates of errors.

\noindent To better understand these differences  I applied my own mode fitting code (described below)  to 16CygA\&B and KIC 6116408, 8379927 and 10454113, using the kasoc power spectra, Davies's spectra  and my own power spectra derived from kasoc light curves. 
% exploring the effects of different assumptions/constraints on mode height ratios $h_\ell/h_0$, rotation $(\omega, i)$, low signal to noise and different power spectra from the same light curves. 
 I find significant differences between frequencies derived from different power spectra and a smaller difference between weighted and unweighted spectra.
 I find that different mode height ratios $h_\ell/h_0$ (fixed or free) have little effect except for low values;
 % I find that low mode height ratios can give important differences in frequencies but otherwise 
 % differences in ratios (fixed or free)   are not significant, 
 that too low an inclination angle can have a significant effect; and that rejecting modes with low signal to noise  gives very much better agreement between different determinations of frequencies.. Details are presented below.  
  
For 16CygA\&B  I find modest agreement between Davies's ($\nu_D$) and my frequencies ($\nu_{RD}$) using Davies's power spectra  ($\chi_A^2=0.33, \chi_B^2=0.21$) and  very good agreement if I reject modes with low signal/noise 
($\chi_A^2=0.06, \chi_B^2=0.03$). 
I do not find such agreement between the Legacy frequencies ($\nu_{L}$) and my frequencies ($\nu_{RL}$) derived from the  Legacy power spectrum for 16CygA ($\chi^2_A=1.53$). I find much better agreement between my values $\nu_{RD}, \nu_{RL}$  from the two power spectra   ($\chi^2=0.44$).
% I find better agreement with $\nu_{RD}$ from Davies's spectrum ($\chi^2=0.44$). 
 I  show that there are some misfits in the Legacy frequencies for 16CygA (Figure 4 below).

For 16CygB two versions of the power spectrum (v1,v2)  have been listed on the KASOC website,  the earlier version (v1) gives modest agreement between my $\nu_{RL}$ and the legacy values 
($\chi^2_B=0.35$) but these differ substantially from the values $\nu_{RD}$ from Davies's spectrum ($\chi^2=1.04$); using (v2) gives values closer to Davies's values ($\chi^2=0.27$) but a considerably larger  difference from the Legacy values $\nu_L$  ($\chi^2=1.13$). 

For the other 3 stars I find closer (but not good) agreement between my frequencies ($\nu_{RA}$) and Appourchaux's ($\nu_A$) than with the Legacy values ($\nu_L$). 
\footnotesize
\vskip 6pt
1.  A\&A 566, 20A, 2014 (and private communication)~~~~~2. MNRAS, 488, 2959, 2015
\normalsize
\newpage
\large
\centerline {\bf $\chi^2$ of fits of Legacy frequencies to those of Appourchaux  and Davies}
\normalsize
\vskip 6pt
% \centerline{\bf and Davies et al (2015)}
%\vskip 3pt
\noindent $\chi^2_m$ is the fit of Legacy frequencies $\nu_L\pm\sigma_L$  to Davies's $\nu_D\pm\sigma_D$, or  Appourchaux's $\nu_A\pm\sigma_A$, taking $\sigma_m=$max($\sigma_L,\sigma_D$) or max($\sigma_L,\sigma_A$),
  $\chi^2_L$  with $\sigma_L$, $\chi^2_D$  with $\sigma_D$, $ \chi^2_A$  with
$\sigma_A$. $n(\nu)$ is the number of frequencies in common,  $n(<\sigma_m)$ the number of frequencies that agree within $1\sigma_m$. 4 examples are shown below.
\vskip-12pt
\begin{table} [h]
\setlength{\tabcolsep}{13.7pt}
\renewcommand{\arraystretch}{1.0}
%\centerline {\bf Fits of Legacy frequencies to Appourchaux and Davies frequencies}
\vskip 0.15cm
\vskip 5pt
%\tiny
\centering
 \begin{tabular}{c c c c c c c c c c c r c c c   } 

%\hline

%\noalign{\smallskip}
%\tiny
%\small
 Fitter & KIC no & $\Delta$ & $n(\nu)$ & $ n(<\sigma_m)$ & $\chi_m^2$ & $\chi_L^2$ & $\chi_D^2\,/\,\chi_A^2$ ~~~\\ [1ex]
\hline
%\noalign{%%\largeskip}
%\tiny
%\small
Davies &  16CygA   &       103  &       50 &   23 &    1.60 &  1.64  & 11.47\\
Davies &  16CygB   &       117  &       51 &   33 &    1.35  & 1.62 & 1.79\\
Davies &   8379927 &      120 &       49 &       36 &   0.50 &  0. 94   & 0.59 \\
 \noalign{\smallskip}
  \hline 
% \noalign{\smallskip}    
App  &     12317678 &       64 &       52 &       22 &   1.65 &   2.42 &   4.16\\
App  &     12258514 &       75 &       45 &       34 &   0.87 &   1.23 &   1.06\\
App  &     12009504 &       88 &       43 &       30 &   0.85 &   1.08 &   1.25\\
App  &     11081729 &       90 &       40 &       25 &   2.87 &   5.40 &   8.50\\
App  &     10454113 &      105 &       49 &       34 &   1.05 &   1.31 &   1.67\\
App  &     10162436 &       56 &       48 &       26 &   1.53 &   1.83 &   1.88\\
App  &      9812850 &       65 &       48 &       31 &   1.55 &   1.92 &   3.11\\
App  &      9206432 &       85 &       49 &       36 &   0.86 &   0.98 &   1.44\\
App  &      9139163 &       81 &       55 &       39 &   1.51 &   2.63 &   1.90\\
App  &      9139151 &      117 &       34 &       26 &   0.67 &   0.86 &   1.17\\
App  &      8694723 &       75 &       53 &       40 &   1.01 &   1.47 &   1.22\\
App  &      8379927 &      120 &       45 &       35 &   0.71 &   1.15 &   0.73\\
App  &      7206837 &       79 &       43 &       24 &   2.38 &   2.70 &   3.50\\
App  &      7103006 &       60 &       53 &       33 &   2.11 &   2.54 &   3.42\\
App  &      6679371 &       51 &       54 &       33 &   1.21 &   1.83 &   1.82\\
App  &      6508366 &       51 &       50 &       34 &   1.33 &   1.80 &   1.88\\
App  &      6116048 &      101 &       42 &       23 &   1.63 &   2.23 &   1.77\\
App  &      2837475 &       76 &       51 &       33 &   1.10 &   1.56 &   1.69\\
App  &      1435467 &       70 &       45 &       31 &   2.98 &   3.79 &   3.27\\
\noalign{\smallskip}
  \hline
\end{tabular}
 \end{table}
\vskip-10pt
\begin{figure}[h]
\begin{center} 
\vskip 6pt
\includegraphics[ width=17cm,height=11cm]{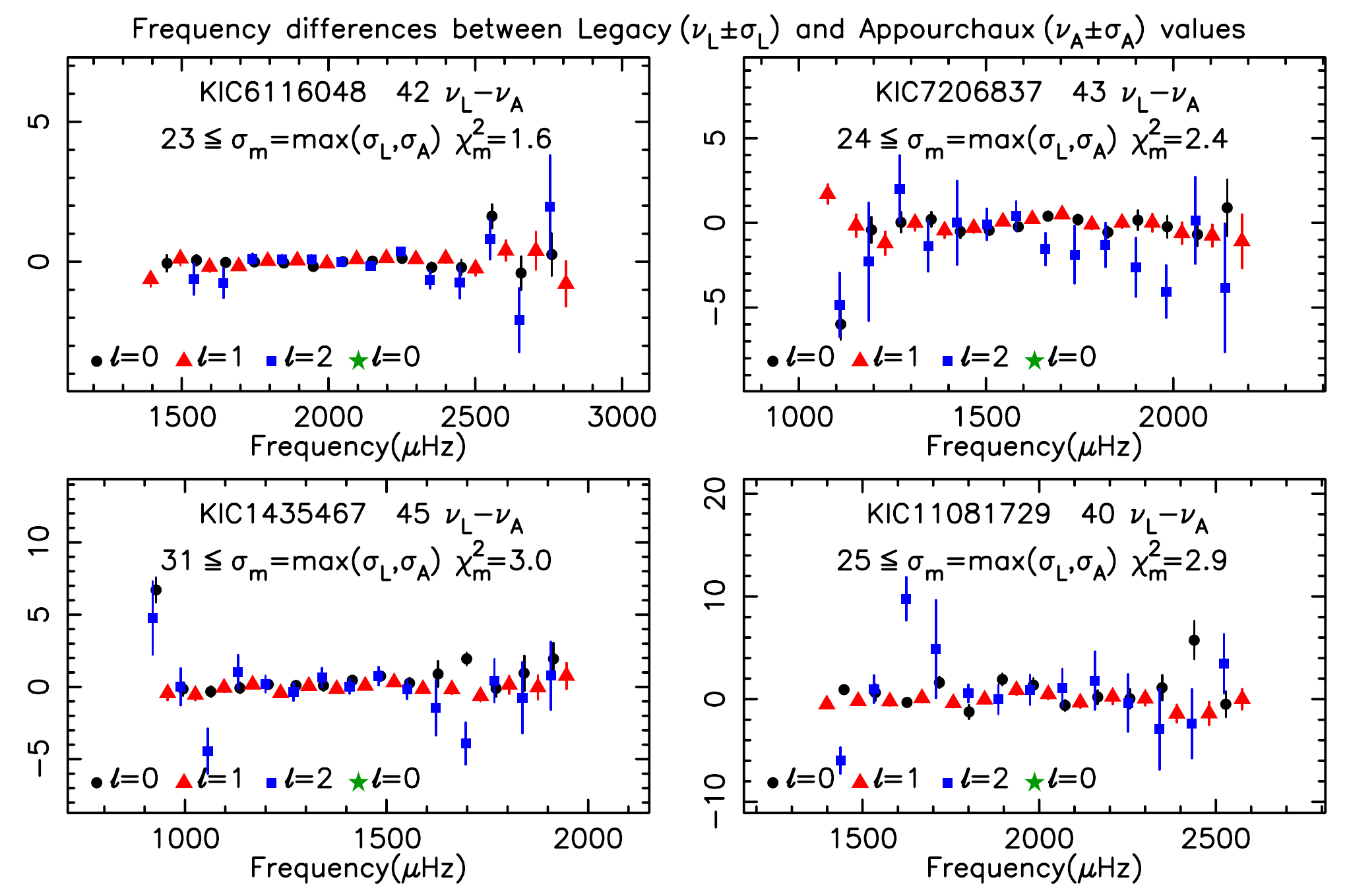}
  \end{center}  
 \end{figure}
 \newpage

\large
\centerline{\bf iwr's frequencies ($\nu_{RL}, \nu_{RD}$)  vs
Legacy ($\nu_L$) and Davies ($\nu_D)$ values for 16CygA\&B}
\normalsize
\vskip 8pt
\noindent {\bf iwr's mode fitting algorithm.}
For given rotational parameters ($\omega, i$)  I search for a global fit of symmetric Lorenzian profiles to a section of the power spectrum that  includes the region of p-mode power, iteratively updating mode pairs to reach a minimum of a maximum likelihood estimator. The background is determined by fitting a Harvey type function to the high and low frequency extremes and to power minus mode power in windows in the central region and updated after each global iteration.. The starting values are given by local fitting of mode pairs. The mode heights and widths for $\ell=1,2,3$ are determined by interpolation in the values for $\ell=0$ with given (or free)  mode height ratios.  This procedure is then repeated with different ($\omega, i$) to find the best fit.
%I seek a global minimum (maximum likelihood) of a fit to a section of the power spectrum that  extends beyond the region of p-mode power. Initial estimates  of the frequencies and ($\omega, i$). are given by individual mode pair fits and the background determined by fitting a Harvey type function to the high and low frequency extremes outside the domain of p-mode power and to power minus mode power in windows in the central region in between the modes. This gives an initial global fit which is iteratively updated replacing mode pairs to improve the global fit taking the height and widths of the $\ell=0$ modes and $\ell=0,1,2,3$ frequencies as adjustable parameters and determining heights and widths of $\ell=1,2,3$ modes by interpolation in the $\ell=0$ values for values with given mode height ratios, and fixed ($\omega, i$). The background is recalculated after each global iteration.  
 %The whole process is repeated until reaching a minimum in the 
%maximum likelihood estimator.  This is then repeated with different  ($\omega, i$) to derive the best fit.
In fitting 16CygA\&B I took Davies's central values for ($\omega, i$)=(0.496, 56) for A and (0.466, 36) for B,  and constant mode height ratios ${\rm  h_1/h_0=1.554,  h_2/h_0=0.582,  h_3/h_0=0.040}$.% these were varied in subsequent runs..
%\normalsize
 \vskip 3pt
\noindent The figures show the differences between my frequencies $\nu_{RD},\nu_{RL}$ from the Davies and KASOC  power spectra for 16CygA\,\&\,B(v2). There is modest agreement with Davies's values for both 16CygA\&B (except for modes with low signal/noise) but not with the Legacy values.
%, and a better fit  to Appourchaux's values ($\nu_A$)  for KIC6116084 than to Legacy values. 
Moreover my frequencies from the 2 power spectra $\nu_{RD},\nu_{RL}$ are in modest agreement with each other ($\chi^2_A=0.44, \chi^2_B=0.27$). This suggests there could be some misfits in the Legacy mode fitting, which is clearly seen for 16CygA on the next page.

\noindent I also find a closer fit  to Appourchaux's values  for KIC6116084, 8379927, 10454113 than to Legacy values.
\vskip 10pt
\begin{figure}[h]
\begin{center} 
\vskip -15pt
   \includegraphics[width=17.5cm,height=16.8cm] {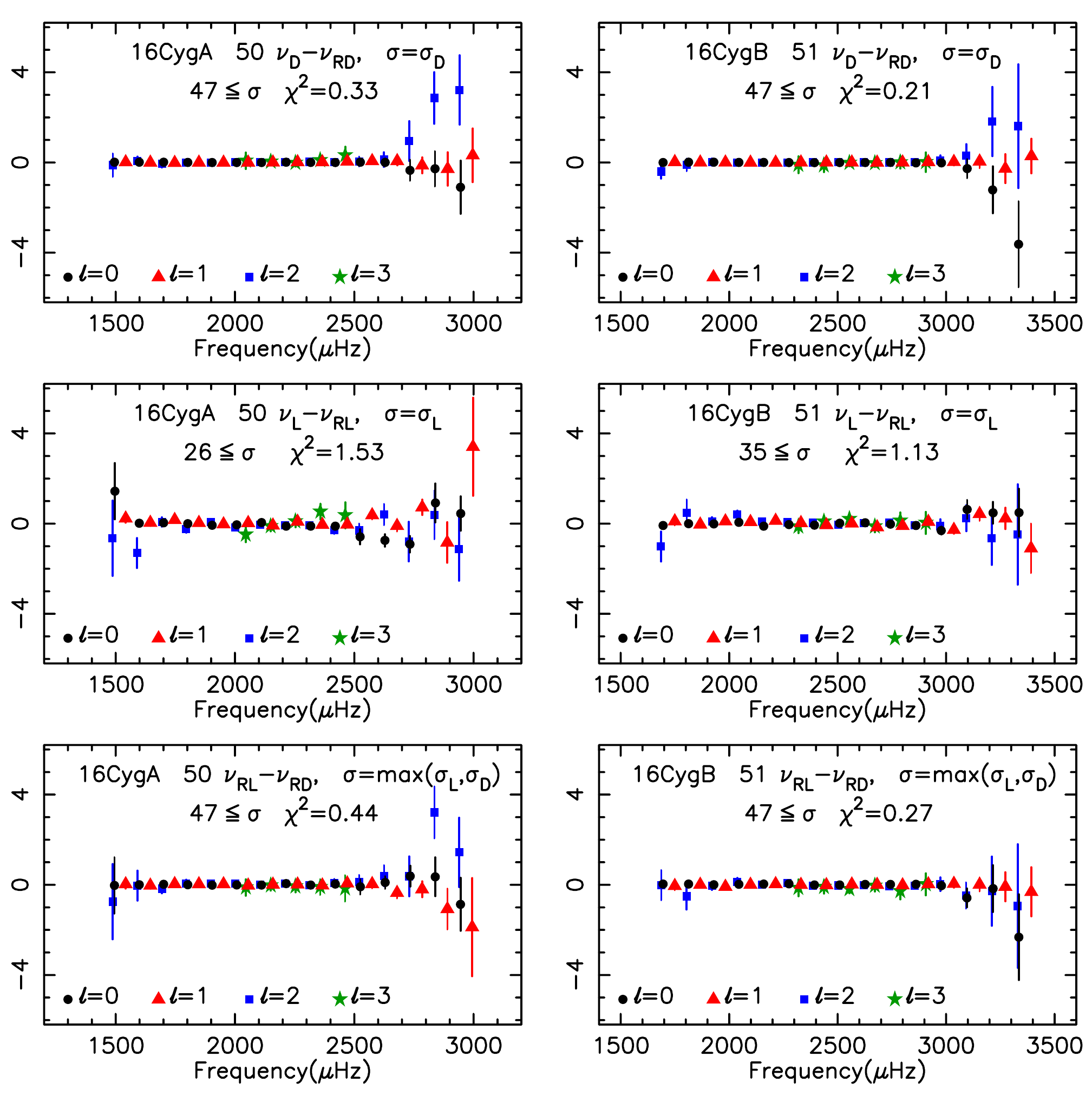}
  \end{center} 
  \end{figure}
%\begin{figure}[h]
%\begin{center} 
%\vskip-25pt
%   \includegraphics[width=17.5cm,height=5.55cm]{freq_diff_6116048.pdf}
%  \end{center}  
%  \end{figure}
\newpage
%%%\large
%vskip 40pt
\large
\centerline{\bf 16CygA:  iwr's fit to the Legacy power spectrum and Legacy frequencies}

\centerline{\bf and fit to Davies power spectrum and Davies frequencies}
\normalsize
\vskip 5pt
The following figure shows (in red) my fit to the full power spectrum used in the Legacy fit (courtesy of M Lund)
  [kplr012069424\_kasoc-wpsd\_slc\_v1.pow] overlaid on a $0.2\mu$Hz boxcar of the spectrum around 3 mode pairs  and and (in blue) the location of the Legacy frequencies, which are not in agreement.  Below is the comparable fit to the Davies power spectrum for 2 of the mode pairs, which are in agreement. This suggests there may be some error in the Legacy fitting algorithm.
\begin{figure}[h] 
\begin{center} 
   \includegraphics[width=17.5 cm]{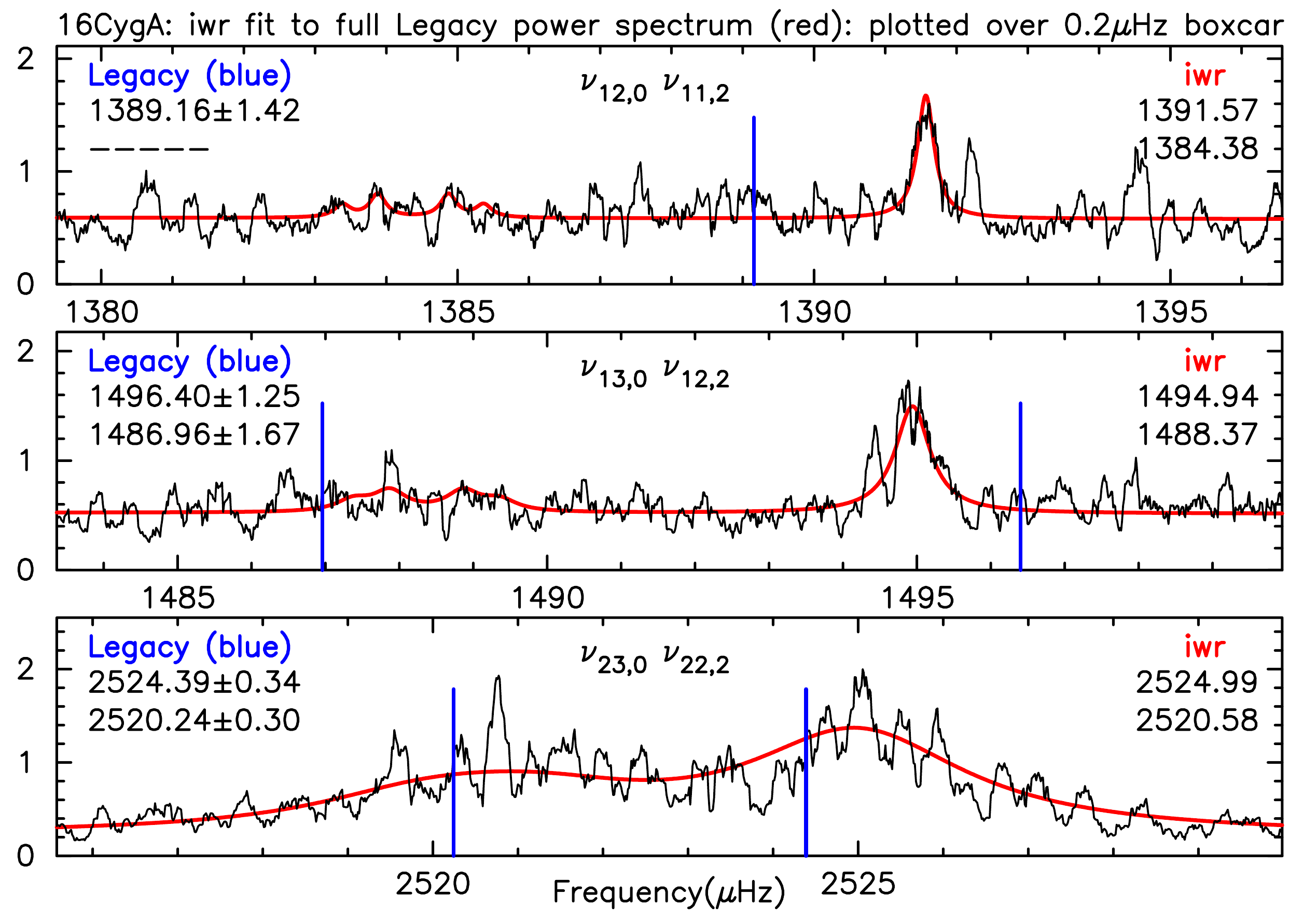}
   \vskip 8pt
    \includegraphics[width=17.5 cm]{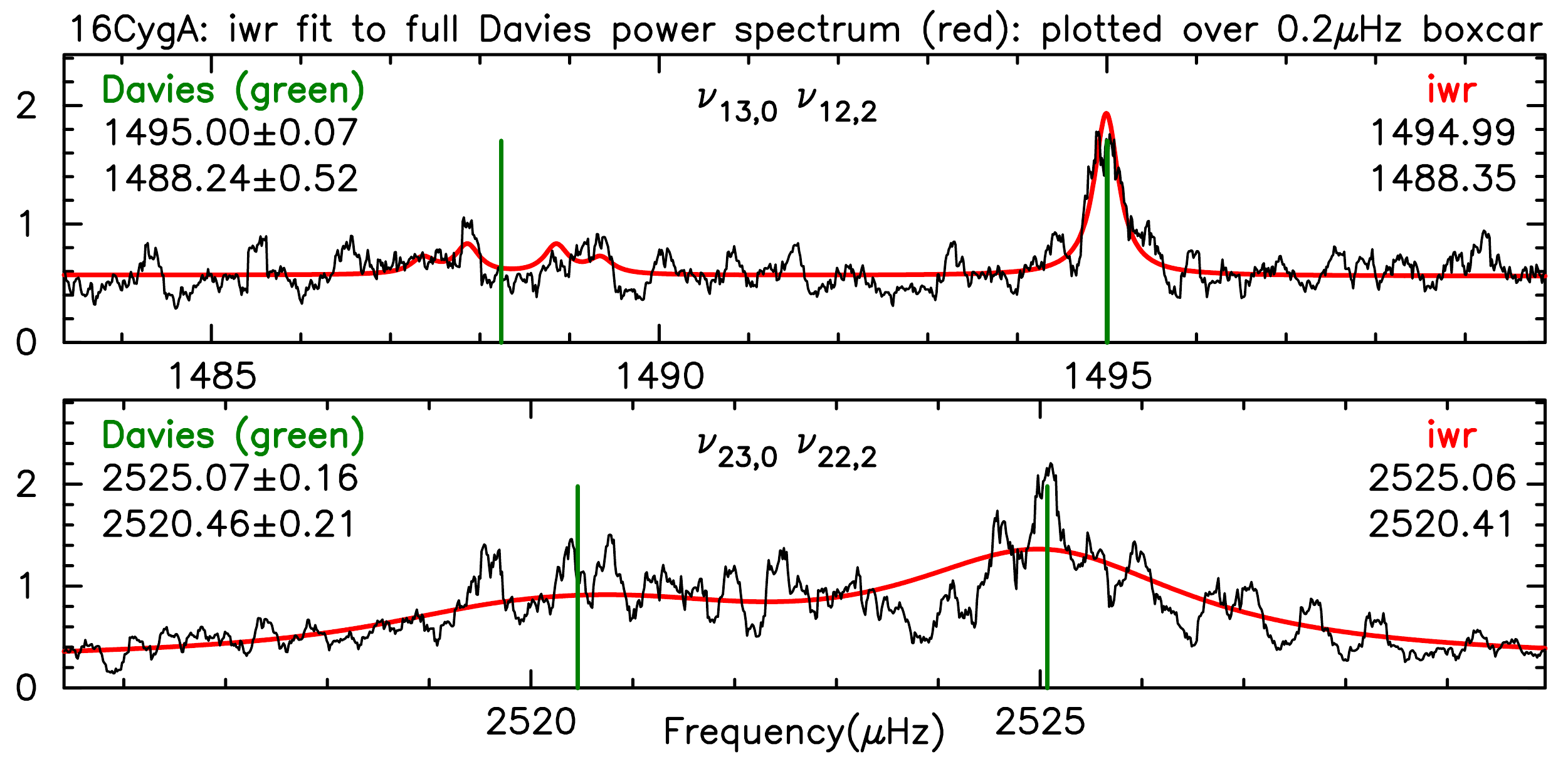}  
    \vskip -20pt
  \end{center}  
  \end{figure}
   \eject
  
\newpage

\large
\centerline{\bf Frequencies  at low signal to noise}
\normalsize
\vskip 5pt
\noindent As shown above I reproduce Davies's frequencies using Davies's power spectra for 16CygA\&B with a 
$\chi_A^2=0.33$ and $\chi_B^2 =0.21$, the major divergences being at low and high frequencies where the mode heights are small compared to the background and so are very sensitive to modelling of the background, and to the derivation of  power spectra from light curves.
% which are done in different ways by different fitters .
This, and large  mode widths at high frequencies, makes me question the reliability of frequency estimates for low signal to noise. 
 
I define signal/noise (S/N) as the maximum height of a (rotationally split) mode divided by the local background; this is shown in the left panels of the following figure for 16CygA\&B for the fits to Davies's power spectra and 16CygB  for the kasoc\_v2 power spectrum; all $\ell=3$ and some  $\ell =0,1,2$ modes have 
 S/N\textless1;  as shown in the  top 2 right panels if these modes  are excluded the quality of the fit of my frequencies 
 ($\nu_{RD}$) to those of Davies ($\nu_D$)  is much improved ($\chi_A^2=0.06, \chi^2_B=0.03$).
 
 The situation is different for 16CygB (and A not shown) using the kasoc\_v2 power spectrum - there is no improvement in the fit of my frequencies $\nu_{RL}$ to the Legacy values ($\nu_L$) - indeed the $\chi^2$ of the fit it is slightly worse than when low S/N modes are included.
 % Since these modes deNote that at low S/N  the values of fitted frequencites  are sensitive to the value of the background  which in turn depends on the paricular algorithm for fitting the background, which is an art rather than a science; and at high frequencies from a contribution from undetected modes. The values for such modes are  therefore suspect. \\
%To illustrate the problems with fitting the Legacy spectra I also show the  results for 16CygA using the kasoc power spectrum - here  the agreement between my frequencies and the legacy frequencies is worse than is is without removing the low S/N modes.
\vskip -4pt
\begin{figure}[h]
\begin{center} 
%\vskip-25pt
   \includegraphics[width=18cm]{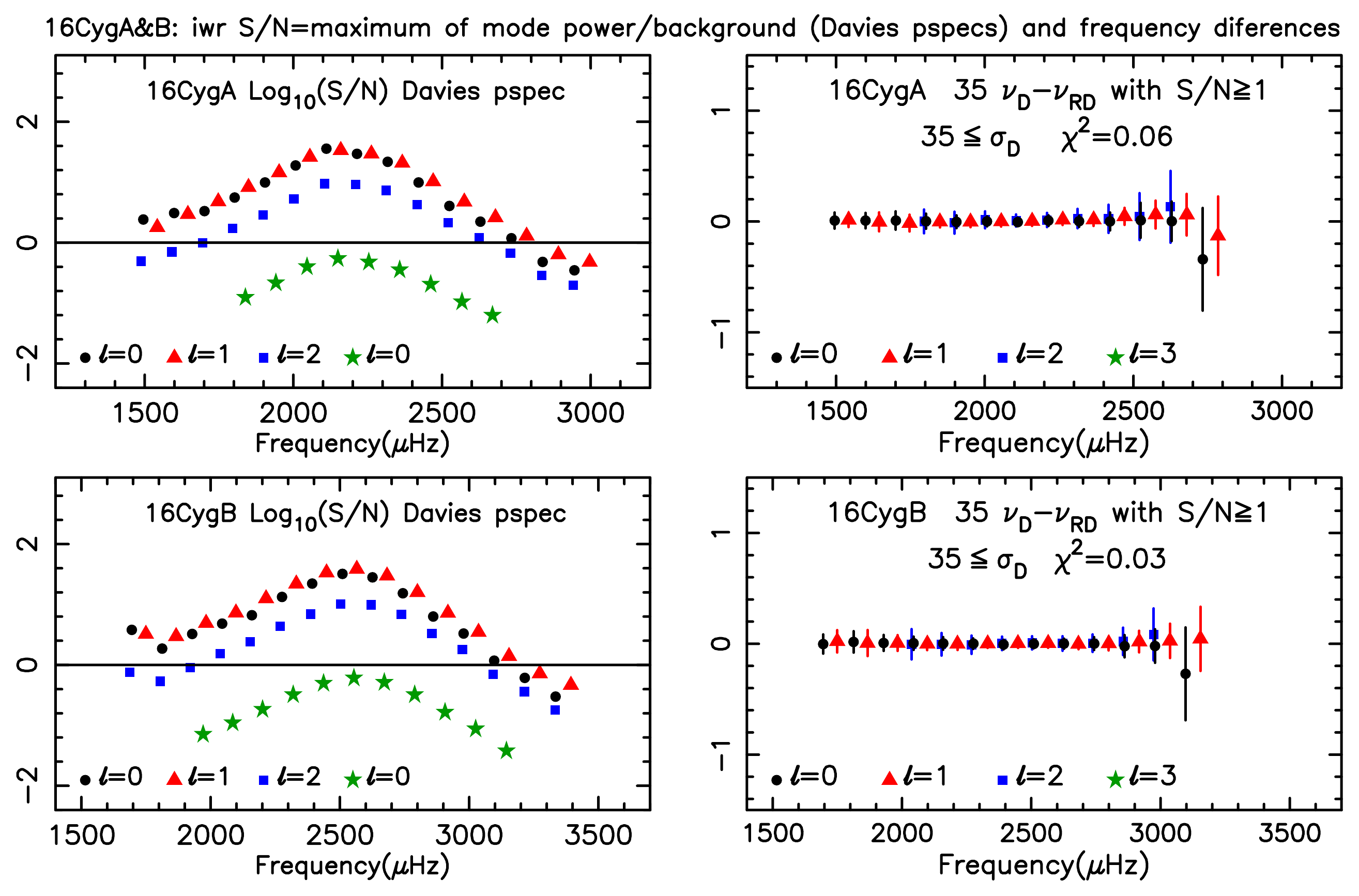}
   \includegraphics[width=17.8cm]{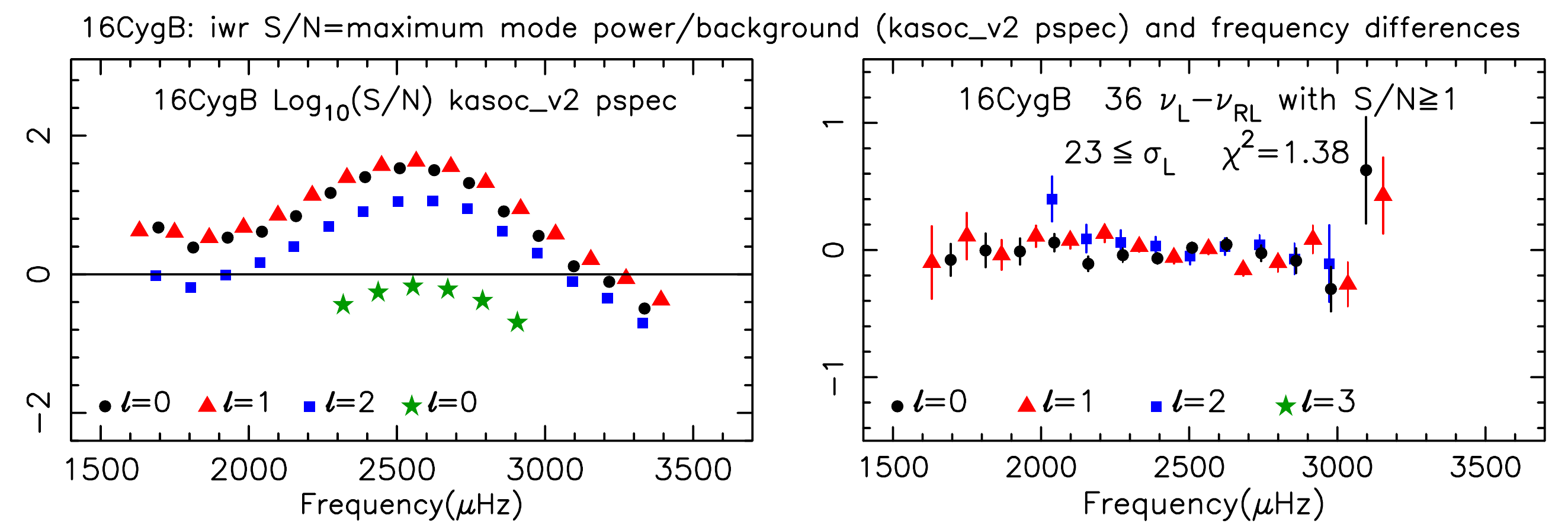}
  \end{center}  
  \end{figure}
 \newpage 

\large
\centerline{\bf Dependence of frequencies on power spectra  derived from kasoc light curves}
\normalsize
\vskip 8pt
%{\bf A: kplr012069424\_kasoc-ts\_slc\_v1.dat~~ B: kplr012069449\_kasoc-ts\_slc\_v2.dat }
The Legacy power spectra for 16CygA\&B were derived from weighted kasoc light curves from Q6-Q17.2 whereas Davies used data from Q7-Q16 with smoothing.  To explore the dependence of the frequencies on power spectra I derived spectra and frequencies $\nu_{DT}$ from Davies's time series (private communication) and 3 sets from the kasoc light curves: $\nu_{12}$ with data from Q6-Q17.2 weighted by the inverse of the flux error; $\nu_{10}$ from Q7-Q16 also weighted, and $\nu_n$ unweighted from Q7-Q16.  I removed bad data (Inf and zero),  had no gap filling, used the Lomb-Scargle algorithm and determined frequencies with the same routine as fits to the Legacy and Davies power spectra ($\nu_{RD}, \nu_{RL} $). The top 2 panels  show near perfect fits of $\nu_{DT}$ to my $\nu_{RD}$ for all frequencies indicating that there is no error in my power spectrum routine.

\noindent The  bottom 4 panels show the best fits of frequencies from the my power spectra ($\nu_{12}, \nu_{10}, \nu_n$) from the kasoc light curves to those from from Davies's and Legacy power spectra taking the Legacy uncertainties on frequencies  $\chi^2$ is for modes with {$S/N\geq 1$} and $\chi^2_f$  for all frequencies.  Not surprisingly the $\nu_{RD}$ best fit $\nu_n$ and $\nu_{12}$  best fit $\nu_{RL}$. 
%The fits of $\nu_{RD}$ to  $\nu_{12}$ has  $\chi_A^2=0.28$ , $\chi^2_B=0.22$ . and
%$\nu_{RL}$ to $\nu_n$ has $\chi_A^2=0.20$, $\chi^2_B=0.22$ , and larger $\chi^2_f$ for the difference for the full frequency set.
The fits of $\nu_{12}$ to $\nu_n$ for S/N$\geq$1 both have $\chi^2=0.21$, and $\sim 0.25-0.5$ for full sets.
%The kasoc light curves taken fr om the web site were:\,
%A: kplr012069424\_kasoc-ts\_slc\_v1.dat (2015-11-29); \,B: kplr012069449\_kasoc-ts\_slc\_v2.dat (2016-05-07 )
%From the kasoc data I removed all entries with zeros or 'Inf', and derived 3 power spectra and corresponding frequencies or both A\&B: 
%$The fits of  $\nu_{RD}$ to $\nu_{10}$ and $\nu_{12}$ show substantial differences with $\chi^2$ ranging from $0.15$ to $1.03$.  The differences between $\nu_{10}$ and $\nu_{12}$ have $\chi_A^2=0.24, \chi_B^2= 0.57$
%and between $\nu_{10}$ and $\nu_n$  have $\chi_A^2=0.06, \chi_B^2= 0.13$

 This suggests that differences in the derivation of power spectra from light curves
  can lead to non-negligible frequency differences in the frequencies.
  %  I stress that the differences are not statistical - given the same input data differences in output are due to different assumptions in deriving  power spectra and frequency fitting.  

 \vskip-4pt 
 \begin{figure}[h]
\begin{center} 
   \includegraphics[width=17.5cm,height=17.6 cm]{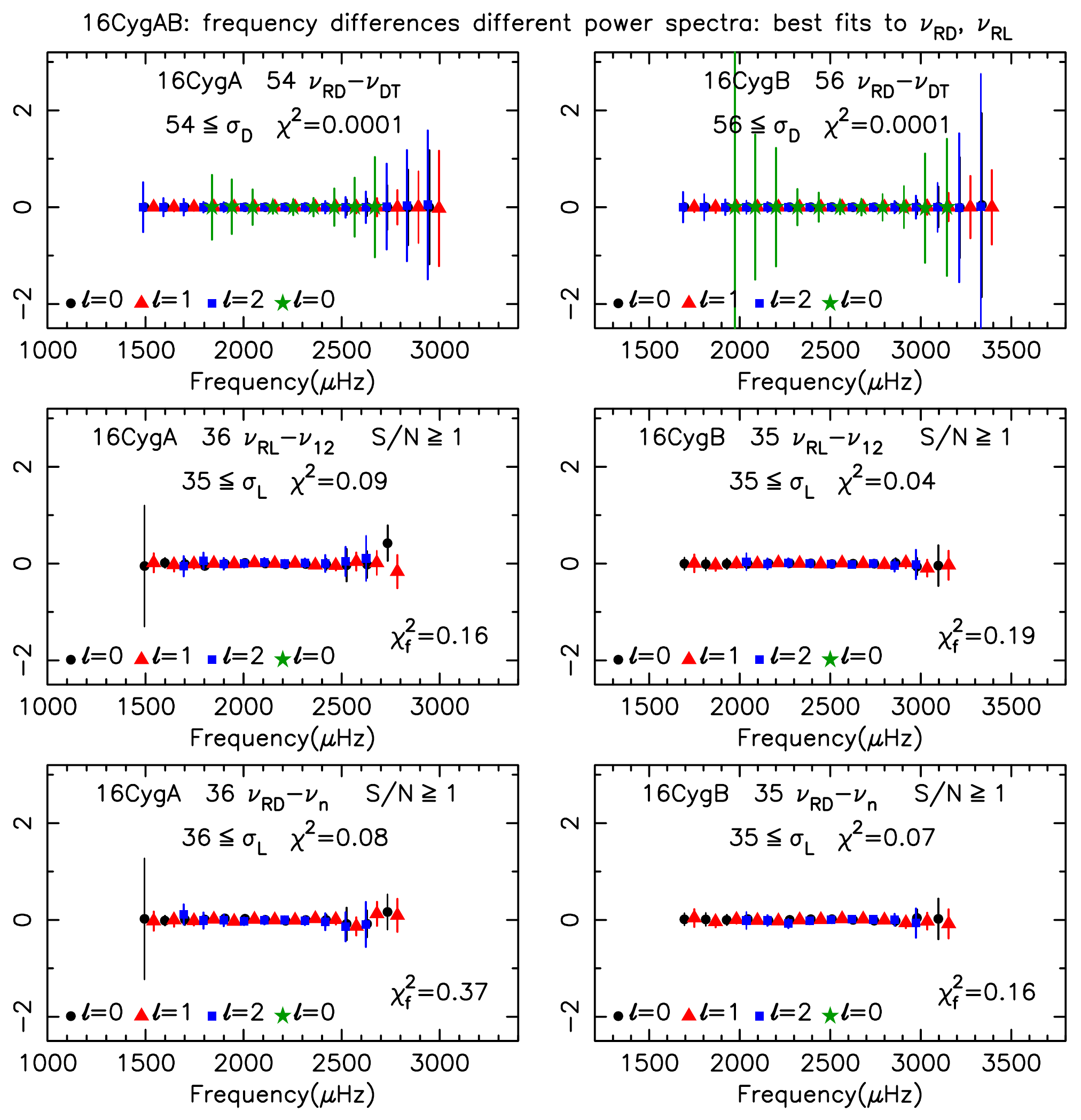}
  \end{center}
  \vskip -20pt
  \end{figure}
\eject
\newpage

  \large
\centerline{\bf Uncertainties  in mode fitting - mode heights}
\normalsize
\vskip 5pt
\noindent As stated above my fits to the power spectra had mode height ratios taken to be constant  and with (my standard) values ${\rm  h_1/h_0=1.554,  h_2/h_0=0.582,  h_3/h_0=0.040}$  which correspond to visibility coefficients for a limb darkening law $f(\mu)= 0.3+ 0.7\mu$ (cf Roxburgh \& Voontsov, 2006)$^1$. I here explore the consequences of alternative fitting models on the results for 16CygB using Davies's power spectrum. I took the following models:
\vskip 4pt
\noindent 1. No $\ell=3$ modes:  The standard values $\rm h_1/h_0=1.554,  h_2/h_0=0.582$ but  setting $\rm h_3=0$
\vskip 2pt
\noindent 2. Highest ratios : $\rm h_1/h_0=1.688,  h_2/h_0=0.800, h_3/h_0=0.109$ , limb darkening law  $f(\mu)= \mu$
\vskip 2pt
\noindent 3. Lowest ratios : $\rm h_1/h_0=1.333,  h_2/h_0=0.313, h_3/h_0=0$ ,  limb darkening law $f(\mu)= 1$.
\vskip 2pt
\noindent 4. Free ratios :  $\rm h_1/h_0,  h_2/h_0, h_3/h_0$ unconstrained but lying between  the values in 2) and 3) above.\\
{\it (Davies et al (2015) take mode heights to be free parameters in their fits to 16CygA\&B}.)
 \vskip 4pt
\noindent The results are displayed in the following diagram which gives the differences in frequencies relative to  my standard fits 
$\nu_{RD}$.  Neglecting the $\ell=3$ modes has a negligible effect on the frequencies but of course decreases the quality of fit of the model to the power spectrum  Only the lowest ratios have a substantial effect on the values of the frequencies, the difference to the standard values decreasing as the height ratios are increased. Letting the ratios be free of course gives the best quality of fit to the power spectrum  since there are more adjustable parameters  - but I question whether this is reasonable.  It may possibly be justified on the grounds that  mode heights are not simply given by limb darkening and may vary with frequency, taking them as free parameters gives some idea of the uncertainties in the frequencies due to uncertainties in mode height ratios. 

In my fitting routine I include $\ell=0,2$ mode pairs  at either end of the range to allow for the contribution of tails outside the frequencies to be determined - changing the constraints on these end values has negligible effect of the values of the frequencies of the modes whose values are sought,
\vskip-5pt
\begin{figure}[h]
\begin{center} 
%\vskip-25pt
   \includegraphics[width=17.4cm]{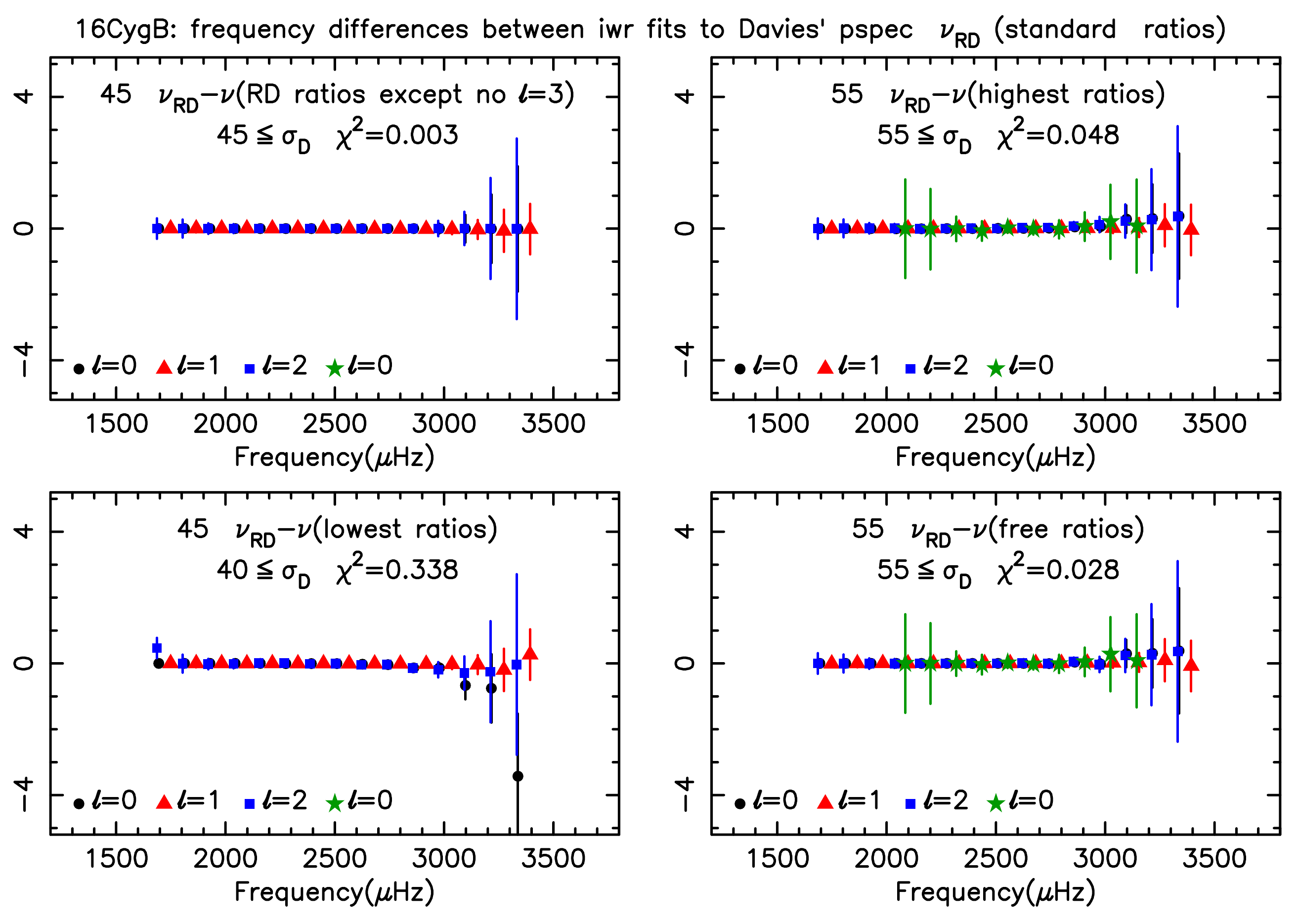}
  \end{center}  
  \end{figure}
\footnotesize
\vskip -25pt
1.  MNRAS, 369, 1491, 2006
\normalsize

\newpage

 \large
\centerline{\bf Uncertainties in rotation}
\normalsize
\vskip 5pt
\noindent  The fits to Davies's power spectra for 16CygA\&B  ($\nu_{RD}$) had rotation and inclination parameters taken as the central values reported in Davies et al (2015): $(\omega, i)=(0.495, 56)$ for A, and $(0.466, 36)$ for B, corresponding to $\omega\sin i=0.411$ for A and $0.274$ for B (my standard values). Here I examine how the value of frequencies depends on the estimation of ($\omega, i)$.
The figure below compares the frequencies of  3 fits to Davies's power spectrum for 16CygA  to $\nu_{RD}$.  All fits had my standard 
fixed mode height ratios corresponding to a limb darkening law $f(\mu)=0.3+0.7\mu$.
I give 3 examples:
\vskip 3pt 
\noindent 1) pole on and/or no rotation;  i=0\\
2) equator on with standard $\omega\sin i = 0.411$, i=90\\
3) my best fit values
\vskip 3pt
The top 2 panels in the figure show the frequency differences and $\chi^2$ of the fits to my reference values $\nu_{RD}$. The pole on case  is a poor fit ($\chi^2=0.51$), the equator on case is better.  The 3rd panel shows the difference between the $i=0$ and $i=90$ cases which gives some idea of the variation in frequencies with assumed inclination.   A more detailed analysis gave  $\chi^2= 0.5, 0.8, 0.6, 0.1$ for $i=10, 20, 30, 40$  all with the reference value of $\omega\sin i = 0.411$
The final panel compares the  best fit values given by searching in a 2-dimensional mesh $(\omega\sin i, i)$  to find the minimum.  The values of $(\omega, i)=(0.494, 52)$ are compatible with those of Davies in spite of the fact that $\omega\sin i=0.388$ is outside their estimated error bars ($\omega\sin i=0..411\pm 0.013$). The difference in frequencies is negligible; $\chi^2=0.004$.
 
Davies et al took free mode height ratios so I repeated the analysis with free ratios; in this case the best fit had 
$(\omega, i)=(0.508, 52)$,  $\omega\sin i=0.400$, $\chi^2=0.003$, but somewhat  larger differences at low $i$

 I did the same for 16CygB  obtaining $(\omega, i)= (0.339, 48)$, $\omega\sin i=0.252$, $\chi^2=0.024$ with fixed height ratios; and 
  $(\omega, i)= (0.346, 50)$, $\omega\sin i=0.265$, $\chi^2=0.048$ with free height ratios.  Here the difference with Davies's values is larger but still the difference in  splitting is very small.  I note that my best fits for A\&B have almost the same inclinations $i=50\pm 2^o$.

  \begin{figure}[h]
\begin{center} 
%\vskip-25pt
   \includegraphics[width=18cm,height=12.5cm] {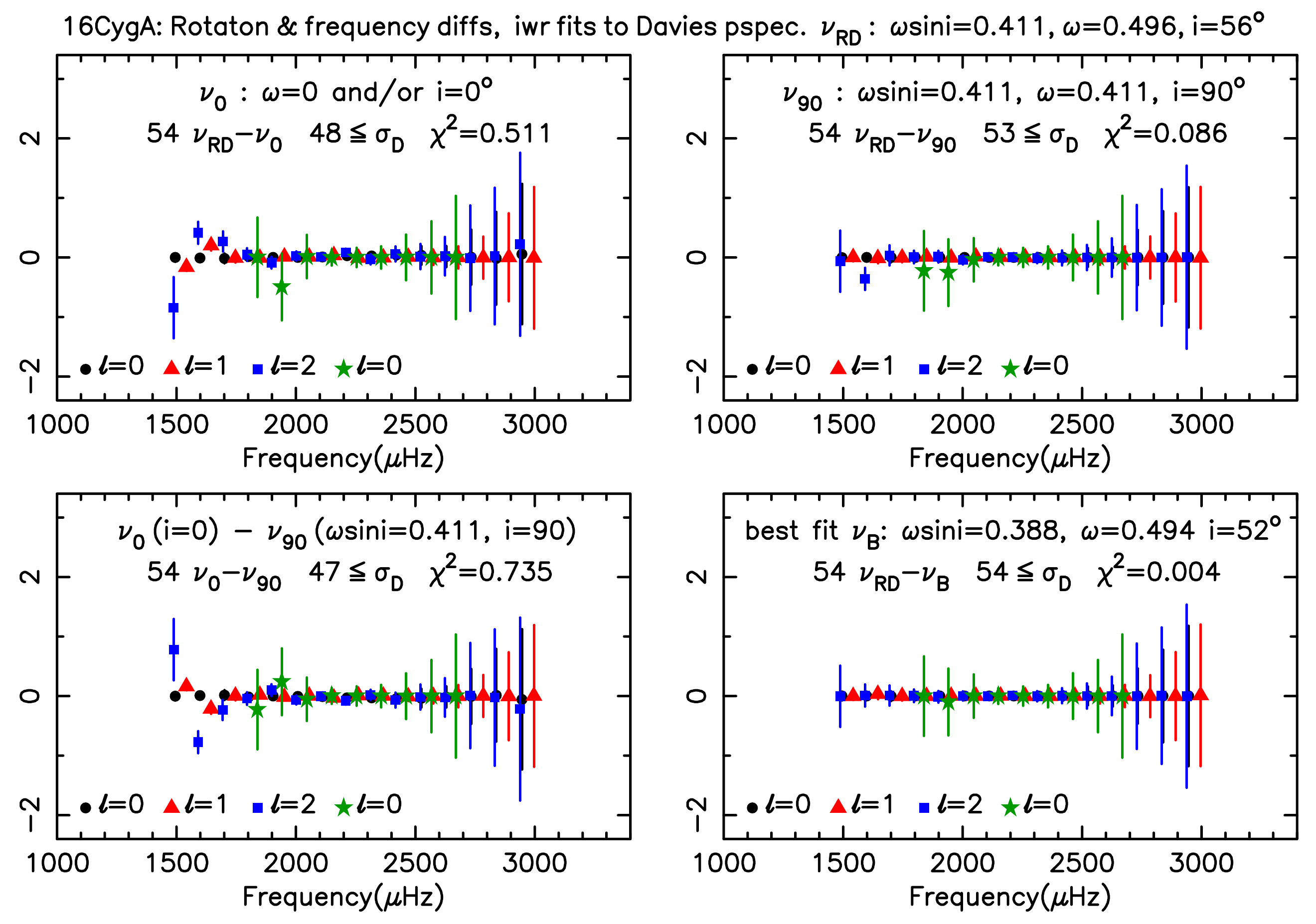}
  \end{center}  
  \end{figure}

% \begin{figure}[h]
%\begin{center} 
%vskip-35pt
%\hskip-15pt
%\includegraphics[width=17.3cm,height=6.2cm]  {MLEvsOT.pdf}
%  \end{center}
 % \end{figure}
%

\end{document}